**Complex magnetic behaviour and evidence of superspin glass state in a binary intermetallic compound $Er_5Pd_2$**

Mohit K. Sharma, Kavita Yadav and K. Mukherjee*

School of Basic Sciences, Indian institute of Technology, Mandi, Himachal Pradesh -175005, India

Email: kaustav@iitmandi.ac.in

**Abstract**

A binary intermetallic compound $Er_5Pd_2$ has been investigated using dc and ac magnetic susceptibilities, magnetic memory effect, isothermal magnetization, non-linear dc susceptibility, heat capacity and magnetocaloric effect studies. Interestingly, even though the compound does not show geometrical frustration it undergoes glassy magnetic phase transition below 17.2 K. Investigation of dc magnetisation and heat capacity data divulged absence of long-ranged magnetic ordering. Through magnetic memory effect, time dependent magnetization and ac susceptibility studies it was revealed that the compound undergoes glass-like freezing below 17.2 K. Analysis of frequency dependence of this transition temperature through scaling and Arrhenius law; along with Mydosh parameter indicate, that the dynamics in $Er_5Pd_2$ is due to the presence of strongly interacting superspins rather than individual spins. This phase transition was further investigated by non-linear dc susceptibility and was characterized by static critical exponents γ and δ. Our results indicate that this compound show signature of superspin glass at low temperature. Additionally, both conventional and inverse magnetocaloric effect was observed with large value of magnetic entropy change and relative cooling power. Our results suggest that $Er_5Pd_2$ can be classified as a superspin glass system with large magnetocaloric effect.

# 1. Introduction

Magnetic materials with competing interactions are being investigated for past several decades because of the exotic magnetic phases exhibited by them. Such materials also exhibit frustrated magnetic behavior, investigation of which is of great interest to the physicists working in this area [1-3]. Such frustrations in magnetic exchange interactions results in a glass-like magnetic state which arise out of cooperative interactions among spins. The prerequisite for the observation of cooperative glassy state is the presence of crystallographic disorder or a geometrically frustrated lattice which generally frustrates the magnetic moment of a magnetic system. This prevents the stabilization of long range ordered ground state and results in spin glass (SG)-like behavior [4]. Inspite of absence of conventional long-range order, these systems exhibit a new kind of order in which the spins are aligned in random directions below the glass transition temperature. It is generally expected that in stoichiometric ordered compounds, SG behaviors will be absent [5]. However, there are recent literature reports about the observation of SG phase in crystallographically well-ordered and stoichiometric compounds like $PrIr_2B_2$ [6], $PrAu_2Si_2$[7], $PrRhSn_3$[8], which is quite exciting from the view point of physics of the SG phenomenon. Khan et al., reported a glassy magnetic state in CrAlGe arising not due to structural disorder, but due to the disorder in the atomic sites of the neighboring atoms [9]. In canonical spin glasses the paramagnetic SG transition occurs due to the presence of disorder or mixed exchange interactions which give rise to atomistic glassy phase with frozen spins. Recent literatures reports [10, 11, 12, and 13] on many magnetic materials suggest that group of spins instead of single spin are frozen below the glass transition temperature and exhibits behavior which is in analogy with that of a prototypical SG. These groups of spins are known as superspin and can have antiferromagnetic, ferromagnetic or ferrimagnetic order with it.

$Er_5Pd_2$ belongs to a family of binary intermetallics $R_5Pd_2$ (R= rare-earth ions) which are known to be good magnetocaloric materials and shows a complex magnetic state. The members of $R_5Pd_2$ family crystallize in cubic structure and have a layered type of crystal structure [14]. Geometrical frustration is negligible in these compounds and there are no reports about the presence of crystallographic disorder. Samantha et al., reported that $Ho_5Pd_2$ have a good magnetocaloric properties arising out of metamagnetic transition below antiferromagnetic ordering temperature [15]. Large change in the magnetic entropy value was also reported for $Dy_5Pd_2$ [16]. Literature reports also suggest the presence of glassy magnetic state in $Tb_5Pd_2$ and $Ho_5Pd_2$ [14]. Hence these compounds can be categorized to the family where origin of glassy behaviour is not due to geometrical frustration or structural disorder. Klimczak et al., reported that the compound $Er_5Pd_2$ orders magnetically and shows spin reorientation [17] whereas Gubkin et al., suggested that the members of $R_5Pd_2$ family may have a glassy magnetic state at low temperature [14]. However, a detailed study about the compound $Er_5Pd_2$ is lacking in literature.

Hence, in this manuscript, we have investigated the magnetic phase of $Er_5Pd_2$ using dc and ac magnetic susceptibilities, magnetic memory effect, isothermal magnetization, non-linear dc susceptibility, heat capacity and magnetocaloric effect studies. Our results proves that, unlike to that reported in literature [17], long range magnetic ordering was absent in this compound. This compound undergoes glass-like freezing below 17.2 K and show superspin glass behaviour. The key observations include: a) a sharp cusp in dc susceptibility along with irreversible behaviour between zero field cooled and field cooled curves, b) field cooled curves increases with decreasing temperature while the irreversibility temperature is suppressed with magnetic field, c) no peak in the heat capacity data d) observation of magnetic memory effect under zero field cooled and field cooled conditions e) a frequency dependent sharp peak in ac susceptibility, f) non-linear dc susceptibility near the glass transition temperature, g) significant value of magnetocaloric effect and relative cooling power. The frequency dependence of the transition temperature was analyzed through scaling and Arrhenius law along with Mydosh parameter while the phase transition was further characterized by static critical exponents $\gamma$ and $\delta$. All our outcomes indicate that the dynamics in $Er_5Pd_2$ is due to the presence of strongly interacting superspins rather than individual spins. In absence of geometric frustration the observed glassiness can be ascribed to the atomic disorder in the compound. Our results also suggest that $Er_5Pd_2$ can be classified as a superspin glass system with large magnetocaloric effect.

**2. Experimental Details**

Polycrystalline $Er_5Pd_2$ has been synthesized by arc melting techniques using stoichiometric ratio of Er and Pd (>99.99 % purity). The pieces were melted several times and a chunk was obtained. For getting better homogeneity the chunk was annealed in evacuated quartz tube at 600 °C for 100 h. X-ray diffraction (XRD) was performed at room temperature in the range 20°-70° with the equal steps of 0.02 using Rigaku diffraction with Cu K$\alpha$ ($\lambda$= 1.54). Figure 1 shows the room temperature indexed X-ray diffraction (XRD) pattern of $Er_5Pd_2$ compound. The compound crystallizes in cubic structure and the obtained lattice parameter was 13.581 Å which was in accordance with literature report [17].Temperature and magnetic field dependent, DC and AC magnetization measurements were performed using Magnetic Properties Measurements System while the temperature dependent heat capacity measurement was performed using Physical Properties Measurements System, both from Quantum Design, USA.

## 3. Results and Discussions

### 3.1 DC Magnetization study

The temperature response of DC magnetic susceptibility ($\chi=M/H$) was obtained under zero field cooled (ZFC) and field cooled (FC) conditions at 100 Oe. Figure 2 (a) shows these curves in the range of 2-300 K. The ZFC plot shows a sharp peak ~17.2K ($T_{ir}$) indicating the presence of a magnetic phase transition which was in accordance to that reported in Ref [17]. A strong irreversibility was observed below 17.2 K between ZFC and FC curves whereas just above 17.2 K the two curves merge with each other. Additionally, FC curve also shows a small peak ~ 17.2 K, followed by an upturn. A peak in the FC curves is generally observed in systems which undergo antiferromagnetic transition while a small peak at glass transition temperature followed by plateau has been reported in the spin glass system [4].Generally, an upturn in the FC curve below the phase transition temperature indicates the presence of clusters of spins [18].Inverse magnetic susceptibility was fitted with Curie-Weiss (CW) law in the range of 20 to 300 K (as shown in inset of figure 2 (a)). The range of fitting of the CW law indicate that spin-clustering was absent above the magnetic phase transition temperature. The effective magnetic moment ($\mu_{eff}$) was estimated from the slope and was found to be around 10.2 $\mu_B$/Er atom which was near the theoretical value (9.6 $\mu_B$/Er atom). The value of CW temperature ($\theta_p$) obtained from fitting was around 21 K. The positive value of $\theta_p$ indicates the dominance of ferromagnetic interactions. This dominance is due to weak polarization of 4d bands of palladium by Erbium, as it has low spin value compared to Dy and Tb [17]. The frustration parameter ($f = \theta_p/T_{ir}$) estimated for this compound was found to be 1.2, implying that geometric frustration is absent [19].

Isothermal magnetization versus magnetic field curve was recorded in the temperature range of 2-100 K and the selected curves are displayed in Figure 2 (b). It is observed that the magnetization does not saturate even under the field of 70 kOe and the curves shows magnetic hysteresis at low temperatures. Similar kind of $M(H)$ curves with non-saturating behaviour had been also observed in other compound showing glassy magnetic behaviour and also in canted AFM systems [4, 20].The observed features in temperature and field magnetization described above has been reported for phase separated systems where ferromagnetic and antiferromagnetic phases co-exists, in glass-like magnetic state, highly anisotropic ferromagnets, in compounds with canted magnetic structure arising from competing AFM and FM exchange interactions and, occasionally, in high-anisotropic antiferromagnets with random orientations of crystallites [21-24]. Thus it appears that below the phase transition temperature the magnetic behaviour of the compound is quite complex.

Hence to shed some light on the magnetic state below 17.2 K, ZFC and FC measurement was done at different applied magnetic field in the range of 0.1 to 10 kOe (Figure 2 (c)).It was noted that with the increase in applied magnetic field, the irreversibility was decreased and it disappeared at 10 kOe. The irreversible temperature ($T_{ir}$) is plotted as a function of magnetic field (as shown in the upper inset of Figure 2 (c)). To check whether this compound shows some signature of glassy magnetic phase, $T_{ir}$ vs. $H$ curve was fitted with mean field theory of SG with random anisotropy [25] by using the following temperature-field phase transition relation:

$$T_{ir}(H) = T_{ir}(0)(1 - AH^p)....(1)$$

where $T_{ir}(0)$ is the transition temperature in the zero magnetic field, $A$ is the anisotropic strength parameter, and $p$ is the exponent. This exponent describes two different regimes; whether the system shows strong anisotropy or weak anisotropy. In the strong anisotropy, *T-H* phase transition line follow the Almeida-Thouluse (*A-T*) line, a line which separates spin glass state of Ising spin glass from its paramagnetic state having $p \sim 0.66$ [26].However, in the weak anisotropy region *T-H* follows the Gabay-Toulouse (*G-T*) line with $p \sim 2$ [27]. For our case, the parameters obtained from fitting were: $T_g = 20.5 \pm 0.7$ K, $A = 0.015 \pm 0.005$ Oe$^{-0.44}$ and $p = 0.44 \pm 0.04$. The value of exponent is not near to that observed for *A-T* and *G-T* lines but it represents a strong irreversibility [28]. This type of behaviour can be due to presence of group of strongly interacting spins rather than single spin. This group of spins along with strong anisotropy of Er ions on application of magnetic field can make the system highly irreversible and can lead to deviation from *A-T* line [29]. For $R_5Pd_2$, $p$ is seen to be different for different members of the family [14], as different rare-earths show different anisotropic behaviour in the applied magnetic field [29]. But, as reported by Wegner and Mydosh [30], such *A-T* phase boundary cannot be considered as an exact proof of glass-like phase transition. However, it should be noted that in this compound no long ranged magnetic ordering was present as was observed from the temperature response of heat capacity which do not show any sharp peak (lower inset of Figure 2 (c)) in the temperature range of measurement. Additionally, it was observed from the figure 2 (b) that with the increase in the temperature the coercive field ($H_C$) sharply decreases. Inset of figure 2 (b) shows the temperature dependent $H_C$ calculated from $M(H)$ isotherms. The curve shows a sharp increment below 10 K. It has been reported in literature that for a single domain particle $H_C$ varies as $T^{1/2}$ [31], while for the canted AFM system, it was found to show parabolic behaviour [20]. Such sharp increment in the value of $H_C$ was noted for systems where the phenomenon freezing of individual spins/group of spins occurs, as was observed in glassy magnetic systems [32, 33]. This curve was fitted using exponential equation of the form [33]

$$H_C(T) = H_C(0)\, e^{-\alpha T} \quad \ldots\ldots(2)$$

where α is the temperature exponent. The obtained fitting parameter are $H_C(0) = 5125 \pm 185$ Oe, $\alpha = 0.47 \pm 0.02$ K$^{-1}$.

### 3.2 Field cooled and zero field cooled memory effect

In order to get a proper insight about this low temperature magnetic phase, measurements of FC and ZFC memory effect was performed by stop and wait protocol [34]. For the investigation of FC memory effect the following protocol was employed: the magnetization data was noted while cooling from 150 to 2 K at 50 Oe applied field. The cooling process was interrupted at 15, 10 and 5 K for waiting time 2 h in each case and during the wait time field was switched off. After the wait time the field was switched on and the cooling data was noted till the next waiting temperature. The curves were plotted in figure 3 (a). A dip was observed in the cooling curve at the wait temperatures which indicates the presence of ageing effect. After reaching 2 K, the sample was immediately heated back without stop upto 300 K, in the same field and at the same rate. The $M(T)$ curve obtained in this way, referred to as $M_{FCW}$, also exhibits peaks at ~5 and 10 K implying that the compound remembers its thermal history or the magnetic state reached during the cooling cycle. To cross-check this feature, immediately after the warming cycle, the compound was again cooled in 50 Oe from room temperature to 2 K without any stop and a field cooled warming measurement was done. In this case, the FC curve [(FC$_{ref}$) in Figure 3 (a)] was smooth with the absence of any step indicating that the observed feature is intrinsic to the compound. Memory effect in the FC magnetization was observed in both superparamagnetic (SPM) and glassy systems [34, 35] whereas, ZFC memory effect is a unique signature of SG behaviour arising from the cooperative spin–spin interaction and is not observed in SPM systems [35, 36]. Hence ZFC memory effect was performed using the following protocol where while cooling the compound in zero field two halt was made for 2 h at 18 and 8 K (shown in Figure 3(b)). The reference curve (ZFC_$_{ref}$) (the normal ZFC curve obtained after cooling to the lowest temperature without any halt) is subtracted from the corresponding heating curve (ZFC_$t_w$) performed after the halts. The difference between these two curves is shown in inset of figure 3(b) which shows a clear dip at the waiting temperature of 8 K confirming the presence of memory effect implying the occurrence of this glassy phase is due to freezing phenomenon. However as expected such feature was absent at 18 K which is above the phase transition temperature of the compound.

### 3.3 Time dependent magnetization study

In order to check whether the low temperature magnetic state is stable or metastable, magnetic relaxation measurement has been performed by using the following protocol: the compound was cooled in zero field from well above the paramagnetic region to measuring temperature (5 and 10 K). Once the measurement temperature was reached, a magnetic field of 50 Oe was applied for 20 min. After that the field was switched off and magnetization was noted as function of time (figure4). It was observed that the magnetization decays with time with the decay rate being more at low temperature implying that the magnetic state is metastable. The data was fitted using the equation

$$M = M_0 - S \ln(1 + t/t_0) \quad \ldots \ldots (3)$$

where $M_0$ is a parameter and $S$ is the magnetic viscosity [4, 5]. Good fits were obtained for all the curves and the fitting parameters for 10 and 5 K are $S = 0.023 \pm 0.002$ emu/gm, $M_0 = 0.36 \pm 0.01$ emu/gm, and $S = 0.014 \pm 0.001$ emu/gm, $M_0 = 0.30 \pm 0.01$ emu/gm respectively. The logarithmic relaxation implies a distribution of energy barrier in the system due to presence of groups of spins of various sizes. Our experimental results suggest the existence of the collective relaxation behaviour in this compound.

### 3.4 AC Susceptibility measurement

The observation of memory effect and non-equilibrium dynamics similar to those observed in spin-glass systems indicates the existence of glass-like phase below 17.2 K. Hence, we had performed ac susceptibility measurements as it is a powerful tool to study the dynamical properties of glass-like system. Figure 5 shows the temperature dependent in-phase ($\chi'$) and out-of phase ($\chi''$) components of ac susceptibility at the different frequencies. The $\chi'$ verses $T$ plot, exhibits a frequency dependent peak around 17.2 K at 13 Hz. This peak temperature was seen to increase with frequency; a common feature observed for glassy magnetic systems. Occurrence of frequency dependent peak in a compound also indicates the absence of long range magnetically ordered state. Hence we ascribe the observed transition at 17.2 K to glass transition. Hence it can be said that $Er_5Pd_2$ undergoes a transition from a disordered phase to frozen state below 17.2 K. Characterization of the frequency dependence of this glass transition temperature($T_f$) is carried out by relative shift in $T_f$ per decade of frequency by Mydosh parameter [4]:

$$\delta T_f = [\Delta T_f/(T_f(\Delta \log f))] \quad .......(4)$$

This parameter has been used to classify various glassy magnetic systems. The value to $\delta T_f$ for this compound was found to be ~ 0.02, which is in between the values reported for non-interacting ideal superparamagnet ($\delta T_f$ ~ 0.27) and canonical SG system ($\delta T_f$ ~ 0.005) [4]. The value is near to that observed for compounds which show superspin glass behaviour [37, 38]. In $\chi''$ an additional broad shoulder was noted below 10 K which signifies that possibly some spin dynamics is being changed. The temperature dependence of the relaxation time $\tau$ can be describe by the power law which is expressed as [4]

$$\tau = \tau_0 (T/T_{f0}-1)^{-zv} .......(5)$$

where z is the dynamic exponent which describes the slowing down of relaxation, ν is the critical exponent which describes the growth of spin correlation length, $T_{f0}$ is the characteristic glass transition temperature and $\tau_0$ is the relaxation time of single spin flip. Inset (a) of Figure 5shows the scaling of $\tau$ with the reduced temperature $\varepsilon = [(T - T_{f0})/T_{f0}]$. The solid line is the fit to the experimental data and a good fit is obtained for the entire frequency range of 13 to 931 Hz. The value of the fitting exponent $zv$ and $T_{f0}$ was found to be ~ 4±0.2 and 17 K respectively.Interestingly the microscopic flipping time $\tau_0$ is found to be $10^{-9}$ sec which is a couple of orders higher than those observed for atomic SG (~ $10^{-13}$s) [4, 5, 8]. The observed values were similar to that noted for systems showing superspin glass behavior [13, 38-39]. This implies that the dynamics in $Er_5Pd_2$ is due to the presence of strongly interacting superspins rather than individual spins. This statement is further substantiated from the deviation from the Arrhenius law [5]

$$\tau = \tau_0 \exp(E_a/k_B T) \quad ....(6)$$

where $\tau_0$ is the characteristic relaxation time, $k_B$ is the Boltzmann constant and $E_a$ is the average thermal activation energy. For single spin flips, the ($1/T$) variation of ln ($\tau$) is linear. But as observed from the right inset (b) of figure 5, this variation is non-linear for Er$_5$Pd$_2$. It indicates that the freezing phenomenon in this compound was of cooperative character. This observation is not unusual and has been reported by Anand et al., in PrRhSn$_3$ compound [8]. Therefore, based on these results it can be said that Er$_5$Pd$_2$ belongs to the category of superspin glass system.

### 3.5 Nonlinear DC magnetic susceptibility

To further characterize this low temperature phase, nonlinear dc susceptibility was calculated since as far as dynamics of phase transition is concerned, this tool is a more powerful technique, as it gives the details of local magnetic behavior. The nonlinear parts of DC magnetic susceptibilities were obtained from the field and temperature dependence of magnetization by using the following expression [39-41]:

$$M/H(T) = \chi_1(T) - \chi_3(T)H^2 + O(H^4) \quad \text{....... (7)}$$

$$= \chi_1(T) - a_3(T)\chi_1^3 H^2 + O(H^4)$$

where $\chi_1$ is the linear susceptibility, $a_3$ (= $\chi_3/(\chi_1)^3$) and the other higher order terms corresponds to the nonlinear contribution. Also, $a_3$ can be used to distinguish between glassy system and a SPM system, as in the latter case, $a_3$ is independent of the temperature, whereas, for the former case, a temperature dependent behaviour is observed [28]. Hence the total nonlinear part of the magnetic susceptibility is expressed as

$$\chi_{nl}(T,H) = 1 - M(T,H)/\chi_1 H \quad \text{....... (8)}$$

For calculate the $\chi_{nl}$, the magnetization data was collected by using the protocol, as described in [39, 41]. Figure 6(a) shows the temperature dependent field cooled magnetic susceptibility ($M/H$) at different applied field. To extract $\chi_1$ and $\chi_3$, this magnetic susceptibility data is plotted with respect to $H^2$ and was fitted with equation (7). The higher order terms were neglected during fitting as at low applied field first two terms were dominant. Temperature response of $\chi_1$ obtained from this exercise was shown by a solid line in figure 6 (a). Inset of Fig 6 (a) shows $\chi_{nl}$ (calculated using equation (8)) plotted as function of $H^2$, at different temperatures. Interestingly, from the figure it is observed the curvature is most significant at 17 K which was the glass transition temperature. For the conventional SG, $\chi_{nl}$ generally follows temperature and field dependent power law behaviour of the form [5,41]

$$\chi_{nl} \sim \tau^{-\gamma} \text{ and}$$
$$\chi_{nl} \sim H^{2/\delta} \quad \ldots\ldots\ldots (9)$$

where γ and δ are two different static critical exponents used for characterizing SG transition. The value of γ was obtained from the logarithm plot between $a_3$ and τ (shown in Figure 6 (b)) while δ was calculated from the linear fitting of ln ($\chi_{nl}$) vs. ln $H$ (inset of Figure 6 (b)), both at 17 K. The obtained value of γ and δ was 2.4±0.1 and 4.1±0.2 respectively. Both the values are similar that expected for a conventional SG [5,41-42]. Hence from the above observations it can be said that in the compound the glassy behavior is introduced by random dipolar interaction among the group of spins and the system undergoes a superspin glass like freezing at low temperature. As geometric frustration is absent in $Er_5Pd_2$ and also there are no reports of crystallographic disorder, the observed glassiness can possibly arise due to the frustration related atomic disorder and layer type crystal structure, similar to that observed for $Tb_5Pd_2$ and $Ho_5Pd_2$ [14].

### 3.6 Magnetocaloric effect

It has been reported in literature that the members of $R_5Pd_2$ family exhibits large magenetocaloric effect (MCE) [15, 16]. Hence MCE was calculated from the $M$ ($H$) isotherms (virgin curves) in the $T$ range of 2 – 80 K using the similar procedure as described in Ref. [43]. Figure 7 show the isothermal magnetic entropy ($\Delta S_M$) as function of temperature at different field change for $Er_5Pd_2$. A conventional MCE had been observed over a wide range of temperature. The maximum value of $\Delta S_M$ was around -19.8 J/kg-K for a field change of 70 kOe. Another parameter required to determine the suitability of a material showing magnetocaloric effect is the relative cooling power (RCP). It is the measure of the amount of heat transfer between the reservoirs (cold and hot) in an ideal refrigeration cycle. The RCP is expressed as

$$RCP = \Delta S_M^{max} * \Delta T_{FWHM} \ldots\ldots (10)$$

where $\Delta S_M^{max}$ is the maximum $\Delta S_M$ and $\Delta T_{FWHM}$ is the full width of half maximum of the peak in $\Delta S_M$. RCP is an important parameter from the view point of application. A large value RCP can be achieved either getting by large $\Delta S_M$ or $\Delta S_M$ being distributed over a wide temperature range or both. Inset of figure 7 shows the applied field change response of RCP which is seen to increase with applied field. The observed value of RCP at maximum field was around 869 J/kg. Such observed large value of $\Delta S_M$ and RCP are comparable with another well-known magnetocaloric materials [44] implying that $Er_5Pd_2$ have a good magnetocaloric properties. It should be noted that for a canonical SG generally there is no large MCE as there is no large entropy change above and below the SG transition. Observation of significant value of $\Delta S_M$ and RCP in this compound indicates that there is ferromagnetic ordering within the superspins. Such large values of these parameters are not unusual and had also been observed in the glassy phase of $Ho_2PtSi_3$ compound [45]. It was also noted that $\Delta S_M$ was asymmetrically distributed around the maximum value which is in accordance to the absence of long range ordering in this compound. Additionally, an inverse MCE was also observed in the low temperature and field (below 15 K and for 30 kOe field change) region. In the glassy phase, this positive value can originate due to rearrangements of the superspin ordering (signature of which was also observed in $\chi''$), indicating non-equilibrium magnetic state at low temperature.

## 4 Conclusions

In summary, we report the results of our investigation of the magnetic properties of $Er_5Pd_2$ through dc and ac magnetic susceptibilities, memory effect, isothermal magnetization, non-linear dc susceptibility, heat capacity and magnetocaloric effect studies. Our investigation reveals that the compound undergoes a magnetic phase transition at around 17.2 K, however, long-ranged magnetic ordering was absent unlike to that reported in literature. This transition was identified to be glass-like and the compound shows characteristics of a superspin glass. This observation is quite interesting as this compound shows no geometric frustration. The observed glassiness probably arises due to the frustration related with atomic disorder and layer type crystal structure. However, further investigation through low temperature microscopic measurements are required to ascertain the real reason for the observation of superspin glass behaviour. The compound also shows large magnetocaloric effect and relative cooling power. Our results suggest that $Er_5Pd_2$ can be classified as a superspin glass system with large magnetocaloric effect.

**Acknowledgements**

The authors acknowledge the experimental facilities of Advanced Materials Research Center (AMRC), IIT Mandi. Financial support from the SERB project EMR/2016/00682 is also acknowledged.

**List of figures**

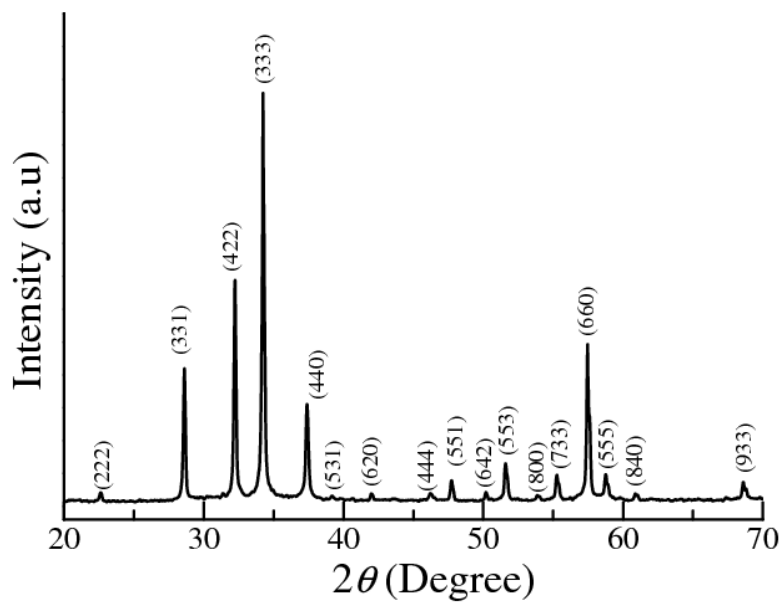

Figure 1 Room temperature indexed XRD pattern of Er$_5$Pd$_2$.

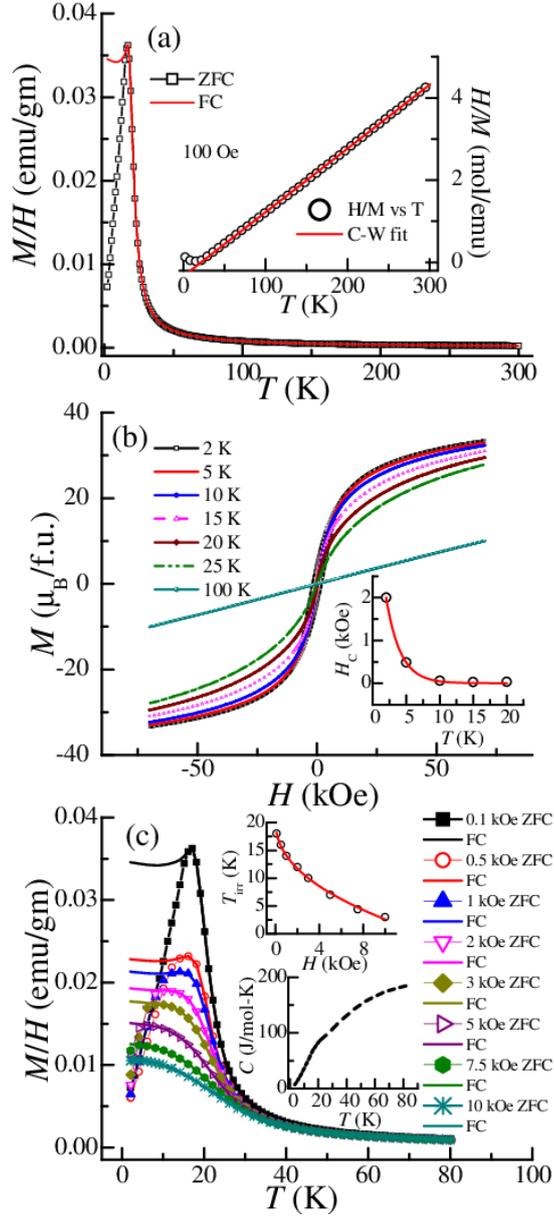

Figure 2 (a) Temperature response of DC magnetic susceptibility ($\chi = M/H$) of Er$_5$Pd$_2$ measured at 100 Oe under ZFC and FC conditions. Inset: Curie Weiss fit of $\chi^{-1}$ vs $T$ at 100 Oe in the temperature range 2-300 K. (b) Magnetic isotherms $M(H)$ at different temperature (2 to 100 K). Inset: Coercive field ($H_C$) vs. temperature plot: Solid red line depicts the exponential function fit using equation (2). (c) Temperature dependent ZFC and FC magnetisation at different applied fields (from 0.1 to 10 kOe) in the temperature range 2-80 K. Upper inset: $T_{irr}$ vs. $H$ plot obtained from ZFC-FC magnetization and fitted (solid red line) with equation (1). Lower inset: Temperature response of heat capacity ($C$) in the temperature range 2-80 K of Er$_5$Pd$_2$.

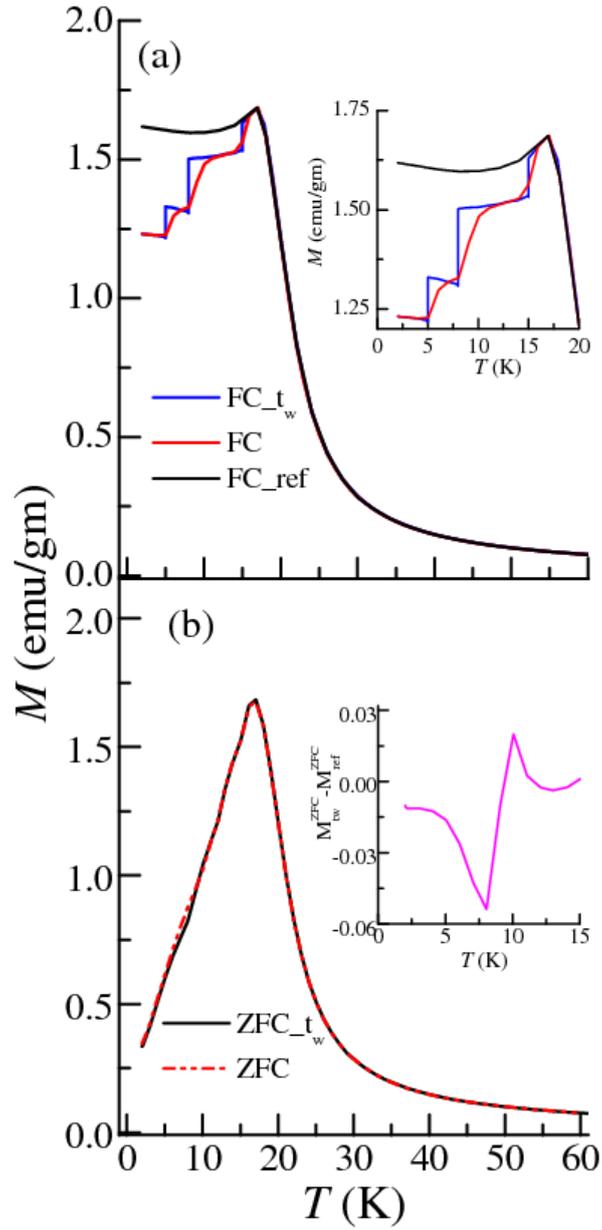

Figure 3 Temperature dependent memory effect of $Er_5Pd_2$ at different waiting temperatures: (a) under FC condition. Inset: Enlarged view of the FC memory curve from 2-20 K. (b) under ZFC condition. Inset: Temperature response of ($M^{ZFC}_{tw} - M^{ZFC}_{ref}$) in the range 2-15 K.

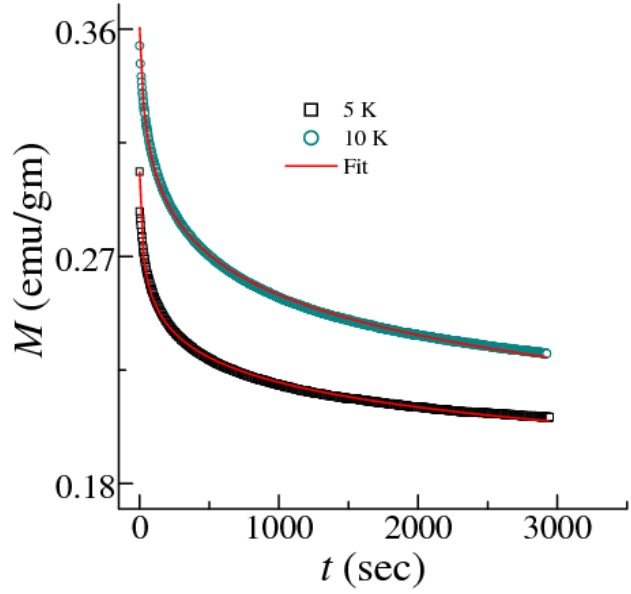

Figure 4 Magnetization as a function of time at 5 and 10 K. Solid red line: Logarithmic fit using equation (3).

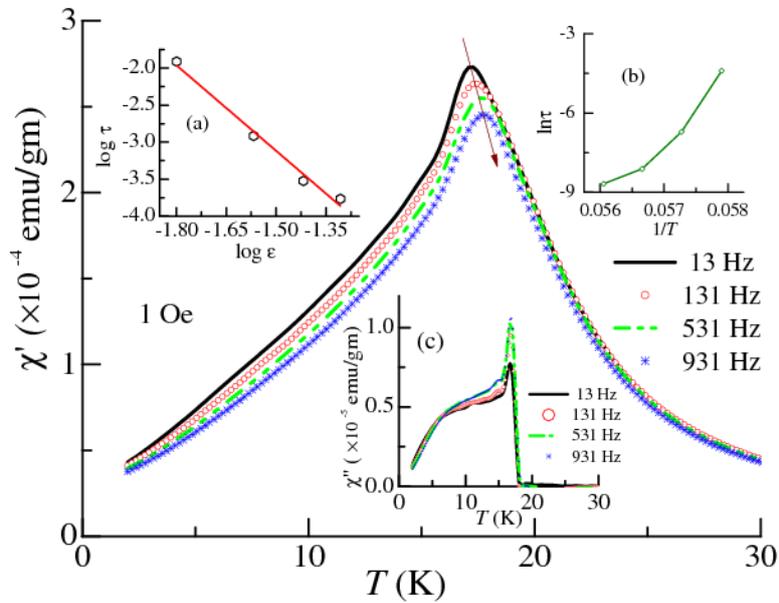

Figure 5 Temperature dependence of the in-phase ($\chi'$) and out-of phase ($\chi''$) (inset (c)) components of AC susceptibility measured at different frequencies (13, 131, 531 and 931 Hz). Inset (a): Power law fit of relaxation time($\tau$) as a function of reduced temperature ($\varepsilon$) using equation (5). Inset (b): Arrhenius fit of relaxation time as a function of inverse temperature using equation (6)

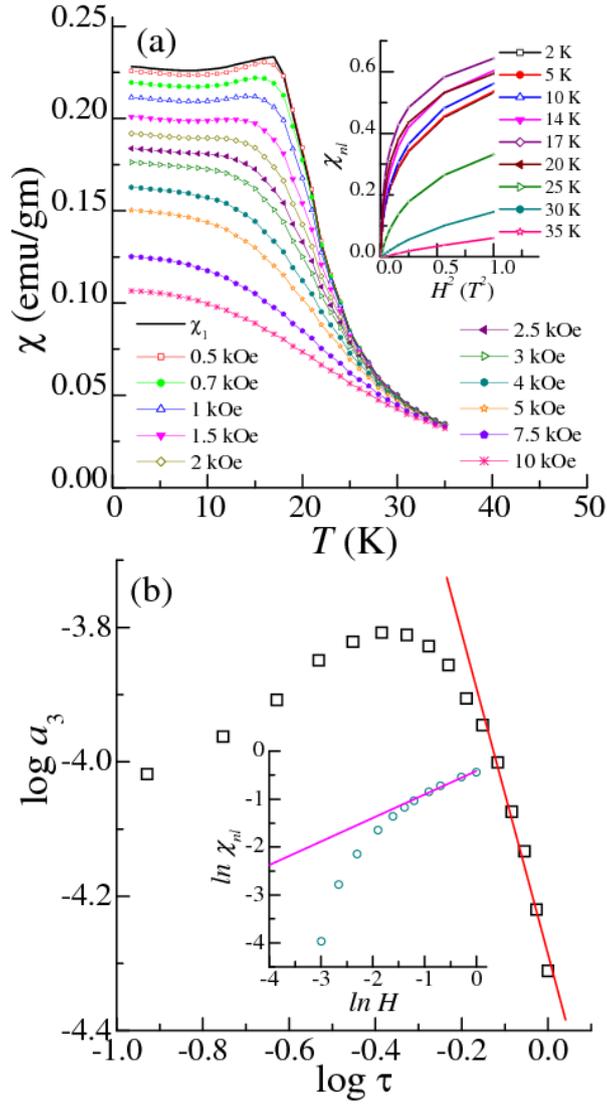

Figure 6 (a) Field cooled magnetic susceptibility at different magnetic fields under slow cooling condition. Solid black line represents linear susceptibility $\chi_1$ obtained by fitting the experimental data using equation (7). Inset: Non-linear susceptibility behaviour ($\chi_{nl}$) as function of $H^2$ for selected temperatures. (b)Plot of log $a_3 (= \chi_3/\chi_1^3)$] as a function of log $\tau$ (=$T/T_0$-1)] at 17 K. Inset: log–log plot of $\chi_{nl}$ vs $H$ corresponding to 17 K.

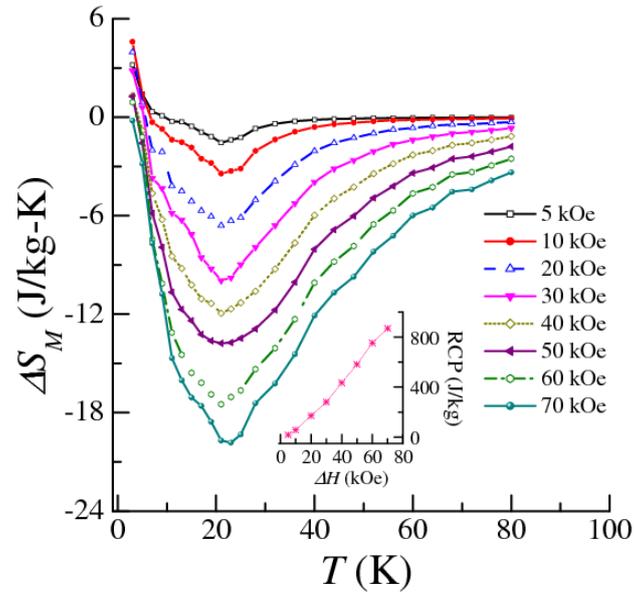

Figure 7 Temperature dependent magnetic entropy change ($\Delta S_M$) under different applied magnetic field change ($\Delta H$ = 5 to 70 kOe) of $Er_5Pd_2$. Inset: Relative cooling power (RCP) plotted as function of $\Delta H$.